\documentclass{aa}
\usepackage{natbib}
\usepackage{url}             

\newcommand{\beq}{\begin{equation}}
\newcommand{\eeq}{\end{equation}}
\newcommand{\beqa}{\begin{eqnarray}}
\newcommand{\eeqa}{\end{eqnarray}}
\newcommand{\beqann}{\begin{eqnarray*}}
\newcommand{\eeqann}{\end{eqnarray*}}
\bibpunct{(}{)}{;}{a}{}{,} 

\begin{document}


\title{
New analytical and numerical models of 
a 
solar coronal 
loop: I. Application to forced vertical kink oscillations 
}

\author{
K.~Murawski\inst{\ref{inst1}}
\and 
A.~Solov'ev\inst{\ref{inst2}}
\and
J.~Kra\'skiewicz\inst{\ref{inst1}}
\and
A.K.~Srivastava\inst{\ref{inst3}}
}

\institute{Group of Astrophysics, University of Maria Curie-Sk{\l}odowska, ul. Radziszewskiego 10, 20-031 Lublin, Poland \label{inst1}
\and 
Central (Pulkovo) Astronomical Observatory, Russian Academy of Sciences, St. Petersburg, Russia \label{inst2}
\and 
Department of Physics, Indian Institute of Technology (Banaras 
			Hindu University), Varanasi-221005, India \label{inst3}
}

\titlerunning{Coronal loop models}
\authorrunning{K.~Murawski et al.}

\abstract
{}
{
We construct a new analytical model of a 
solar coronal loop that 
is embedded in a gravitationally stratified and magnetically confined 
atmosphere. 
On the basis of this analytical model, we devise a numerical model of  
solar coronal loops. We 
adopt it to
perform the
numerical 
simulations of its vertical kink oscillations excited by an external driver. 
}
{
Our model of the solar atmosphere is constructed 
by adopting a realistic temperature distribution 
and specifying the curved magnetic field lines that constitute a coronal loop. 
This loop is described by 2D, 
ideal magnetohydrodynamic equations that are numerically solved by the 
FLASH code. 
}
{
The vertical kink oscillations are excited by a periodic driver in the vertical component of velocity, 
acting at the top of the photosphere. 
For this forced driver with its amplitude $3$ km s$^{-1}$,
the excited oscillations 
exhibit 
about $1.2$ km s$^{-1}$ amplitude in their velocity  
and the loop apex oscillates with its amplitude in displacement of about $100 $ km. 
}
{
The newly devised analytical model of the coronal loops is utilized for the 
numerical simulations of the vertical kink oscillations, which match well with the
recent observations of decay-less kink oscillations excited in solar loops.
The model will have further implications on the study of waves and plasma dynamics
in coronal loops, revealing physics of energy and mass transport mechanisms
in the localized solar atmosphere.
}

%

\maketitle

\section{Introduction}
The solar corona is a magnetically dominated and gravitationally stratified medium 
which can alter the scenario of magnetohydrodynamic (MHD) waves
(e.g., Pascoe 2014). 
Among a number of magnetic structures present there, 
the magnetic loops 
are considered as a major building blocks of the solar corona. 
They are outlined by curved and closed magnetic field lines, which are rooted in the deep atmospheric layers, 
and are built of the denser and hot plasma. 

Coronal loops act as a waveguide for various kinds of magnetohydrodynamic 
(MHD) waves and oscillations. 
Among various modes, the standing, large-amplitude magnetoacoustic kink waves in coronal loops were detected 
in the solar coronal loops (e.g., Aschwanden et al. 1999, Aschwanden et al. 2000, Wang \& Solanki 2004, Wang et al.
2008, Verwichte et al. 2009, Aschwanden \& Schrijver 2011,
White et al. 2012, Srivastava \& Goossens 2013, and references therein). 
These transverse (horizontally and vertically polarized) kink waves are modeled by a number of authors 
(e.g., Gruszecki et al. 2006, Ofman \& Wang 2008, Ofman 2009, Luna et al. 2010, Selwa et al. 2011, Antolin et al. 2014,
and references therein)
who confirmed the observational data by revealing that 
these waves decay on a time-scale comparable to the oscillation wave-period. 
Recently, 
small amplitude transverse waves were 
reported by De Moortel \&
Nakariakov (2012), Nistic\'o et al. (2013) and Anfinogentov
et al. (2013)
who found decay-less oscillations 
with velocity amplitude of few km s$^{-1}$, the displacement amplitude less than $1$ Mm, and wave-periods 
within the range of $2.5$ to $11$ min.

The wave and plasma dynamics of coronal loops highly depend upon their 
plasma and magnetic field  structuring.
The analytical models of coronal loops 
were devised so far for a gravity-free medium and the loop oscillations were 
triggered by impulsive sources in most 
of the theoretical studies. 
Our aim here is to construct 
for the first time an analytical model of a coronal loop in a gravitationally stratified solar atmosphere. 
As the problem is formidable, we limit ourselves to the simplest conceivable case of 
a two-dimensional (2D) model of a coronal arcade loop. 
As a result of very 
long (few thousand lines) analytical expressions which result in the analytical model for 
the equilibrium mass density and a gas pressure, we focus ourselves on the modest case of 
a loop. On the basis of our analytical model, we develop a numerical model of 
a loop. The newly developed coronal loop model will have several applications and studies 
to understand the properties of excited MHD waves and plasma dynamics in such tubes. 
With 
some modification, the model can be adopted to a 
coronal 
loop 
to reveal its wave and dynamical processes. However, as its first application, we study the
vertical kink oscillations evolved into the model loop. These oscillations are excited by 
a forced periodic driver in the vertical component of 
velocity, which acts at the top of the photosphere. 

This paper is organized as follows. The analytical model of a coronal loop 
is introduced in Sect.~\ref{sec:cor_model}.
A numerical model and the
results 
are described in Sect.~\ref{sec:num_sim_MHD}. 
This paper is concluded by a short summary in Sect.~\ref{sec:summary}. 
%
 \section{The analytical model of a coronal loop}\label{sec:cor_model}
%
\subsection{MHD equations}\label{sec:equ_model}
%
We consider a coronal plasma which 
is 
described by 
the ideal magnetohydrodynamic (MHD) equations 
 %
 \beqa
 \label{eq:MHD_rho} 
 {{\partial \varrho}\over {\partial t}}+\nabla \cdot (\varrho{\bf V})=0\, ,\\
 \label{eq:MHD_V}
 \varrho{{\partial {\bf V}}\over {\partial t}}+ \varrho\left({\bf V}\cdot \nabla\right)
 {\bf V}= -\nabla p+ \frac{1}{\mu} (\nabla\times{\bf B})\times{\bf B} +\varrho{\bf g}\, , \\
 \label{eq:MHD_B}
 {{\partial {\bf B}}\over {\partial t}}= \nabla \times ({\bf V}\times {\bf B})\, , 
 \hspace{3mm} 
 \nabla\cdot{\bf B} = 0\, , \\
 \label{eq:MHD_p}
 {\partial p\over \partial t} + {\bf V}\cdot\nabla p = -\gamma p \nabla \cdot {\bf V}\, ,
 \label{eq:MHD_CLAP}
 \hspace{3mm} 
 p = \frac{k_{\rm B}}{m} \varrho T\, ,
 \eeqa
where ${\varrho}$ is mass density, ${\bf V}$ represents the plasma velocity, $p$ is a gas pressure, 
${\bf B}$ is the magnetic field, $T$ is a temperature, $k_{\rm B}$ is Boltzmann's constant, 
$\gamma=5/3$ is the adiabatic index, $m$ is a particle mass, 
that is specified by mean molecular weight of $0.6$, and ${\bf g}=(0,-g,0)$ is the gravitational 
acceleration. 
The value of $g$ is equal to $274$ m s$^{-2}$. 
%
%
\subsection{Equilibrium conditions}
 \label{sec:equil}
We 
assume 
that the above system is 
invariant along the horizontal coordinate, $z$ ($\partial/\partial z = 0$) and 
set 
the $z$-components of the velocity, $V_{\rm z}$, 
and magnetic field, $B_{\rm z}$, to zero.

 \begin{figure}[!ht]
 	\begin{center}
 		\includegraphics[scale=0.5, angle=0]{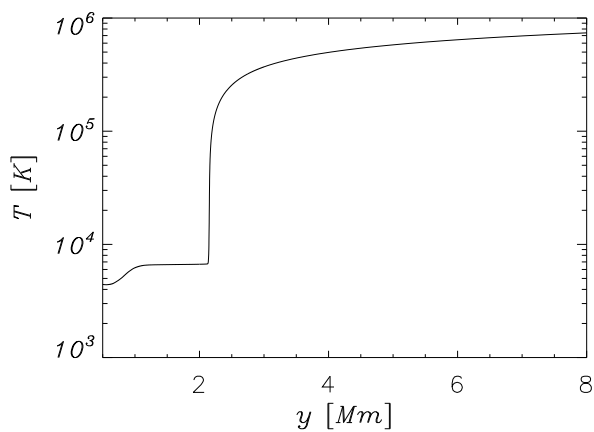}
 		\caption{\small The profile of hydrostatic solar atmospheric temperature vs. hight $y$. 
                         }
 		\label{fig:TTT}
 	\end{center}
 \end{figure}

\begin{figure}[!h]
	\begin{center}
		\includegraphics[width=7.25cm, angle=0]{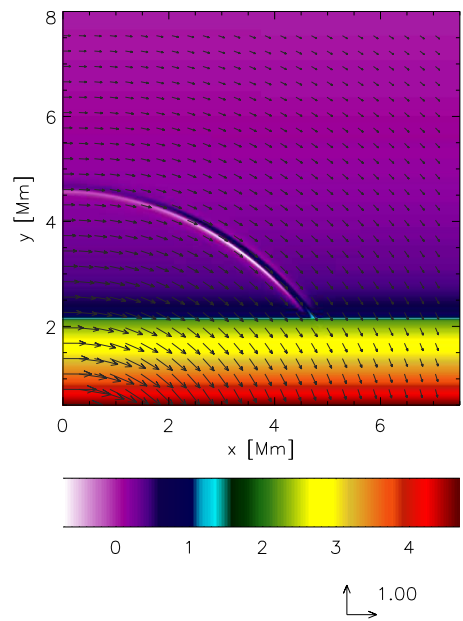}
		\includegraphics[width=7.25cm, angle=0]{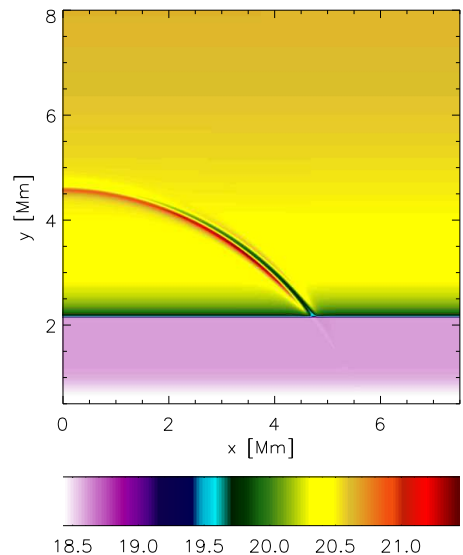}
		\caption{\small 
The top panel: Vectors of equilibrium magnetic field, expressed in units 
                                of $\approx 11.4$ Gauss, and $\log(\varrho)$ (color maps).  
                                The mass density, $\varrho$, 
is given in units of $10^{-15}$ kg m$^{-3}$. 
The bottom panel: Logarithm of temperature (expressed in units of $1$ MK) profile. 
                                The right hand-side of the system (which is symmetric about the vertical $x=0$ axis) 
                                is displayed only. 
                        } 
		\label{fig:mag}
	\end{center}
\end{figure}

In such 
2D
model, the solar atmosphere is in static equilibrium (${\bf V}={\bf 0}$) 
with the Lorentz force balanced by the pressure gradient and gravity forces, and the divergence-free
magnetic field, 
%
\beqa
\label{eq:B}
\frac{1}{\mu}(\nabla \times{\bf B})\times{\bf B} - \nabla p+ \varrho {\bf g} 
= {\bf 0} \, ,\\
\nabla \cdot{\bf B}=0 \, .
\label{eq:solen}
\eeqa
%
%
\subsubsection{Hydrostatic atmosphere}
A hydrostatic atmosphere corresponds to the magnetic-free $({\bf B}={\bf 0})$ case 
in which the gas pressure gradient is balanced by the gravity force 
%
\beq\
\nabla p_{\rm h} = \varrho _{\rm h} {\bf g} \, .
\label{eq:B5}
\eeq
%
With the use of the ideal gas law given by Eq.~(\ref{eq:MHD_p}) 
and the vertical $y$-component of  
Eq.~(\ref{eq:B5}), we express the hydrostatic gas pressure and mass density as
%
\beqa\label{eq:pres} 
p_{\rm h}(y)=p_{\rm ref}~{\rm exp}\left( -
\int_{y_{\rm r}}^{y}\frac{dy^{'}}{\Lambda (y^{'})} \right)\, , 
\hspace{3mm} 
\varrho_{\rm h} (y)=\frac{p_{\rm h} (y)}{g \Lambda (y)}\, ,
\eeqa
%
where
%
\begin{equation} 
\Lambda(y) = \frac{k_{\rm B} T(y)} {mg}\ ,
\end{equation}
%
is the pressure scale-height, and $p_{\rm ref}$ denotes the gas pressure at the 
reference level $y_{\rm r}$ which we set and hold fixed at $y_{\rm r}=10$ Mm.
%


We adopt a realistic plasma temperature profile given by the semi-em\-pi\-ri\-cal 
model of Avrett \& Loeser (2008)
that is extrapolated into the solar corona 
(Fig.~\ref{fig:TTT}). 
In this model, the temperature attains a value of about 
$4300$~K at the bottom of the chromosphere ($y\approx 0.6$ Mm),
$7 \times 10^{3}$~K at the top of the chromosphere ($y\approx 2.0$ Mm). 
At the transition region, which is located at 
$y \simeq 2.1$ Mm, $T$ exhibits an abrupt jump (Fig.~\ref{fig:TTT}), 
and it enhances upto about $0.7 \times 10^{6}$ K in 
the solar corona at $y=8$ Mm. 
Higher in the solar corona, the temperature grows very slowly, tending to its 
asymptotic value of about $1.6$ MK. 
The temperature profile determines uniquely the equilibrium 
mass density and gas pressure profiles which fall-off with the height (not shown here). 
\subsubsection{Magnetic atmosphere}
%
The solenoidal condition of Eq.~(\ref{eq:solen}) is automatically satisfied if we express the equilibrium magnetic field with the 
use of magnetic flux-function ($A(x,y)$) as
\beq
{\bf B}= \nabla \times A {\bf e}_{\rm z}  ,
\label{eq:EqS35}
\eeq
where ${\bf e}_{\rm z}$ is a unit vector along z-direction. 
Setting
\beq
p=p(x,y)=p(y,A) \, ,
\label{eq:EqS5}
\eeq
from Eqs.~(\ref{eq:B}) and (\ref{eq:EqS35}), we get 
\beq
\varrho (y, A)g = - \frac{\partial p(y, A)}{\partial y} \, .
\label{eq:EqS7}
\eeq
Here we infer the hydrostatic condition along the magnetic field line which
is specified by the equation $A = const.$ 
From the $x$- and $y$-components of Eq.~(\ref{eq:B}), 
we obtain the 
equilibrium equation for a system with translational symmetry (Low 1975; Priest 1982): 
\beq
\nabla^2 A = -\mu\frac{\partial p(y,A)}{\partial A} \, ,
\label{eq:EqS11}
\eeq
where $\nabla^2 = \left( \frac{\partial^{2}}{\partial x^{2}} + \frac{\partial^{2}}{\partial y^{2}} \right)$ 
is the Laplacian. 

We assume now that the flux-function $A (x,y)$ 
is known. Therefore, from Eqs.~(\ref{eq:EqS7}) and (\ref{eq:EqS11}), we find the following expressions for $\varrho$ and $p$
(Solov’ev 2010, Kra\'skiewicz et al. 2014, Ku\'zma et al. 2014): 
%
\beqa
\nonumber
\varrho = \varrho_{\rm h} + \\
\frac{1}{\mu g} 
\left[ \frac{\partial}{\partial y} \left( \int \frac{\partial ^{2} A}{\partial y^{2}} 
\frac{\partial A}{\partial x} dx + \frac{1}{2} \left(\frac{\partial A}{\partial x}\right)^{2} \right) -
\frac{\partial A}{\partial y} \nabla ^{2} A \right]   \, ,
\label{eq:B3}
\eeqa
\beq\
p = p_{\rm h} -\frac{1}{2\mu } \left(\frac{\partial A}{\partial x}\right)^{2} 
- \frac{1}{\mu} \int \frac{\partial ^{2} A}{\partial y^{2}} \frac{\partial A}{\partial x} dx \, .
\label{eq:B2}
\eeq
%
%
\subsection{A coronal loop}
For a coronal loop, we make the following choice: 
%
\beqa\nonumber
A(x,y) = S_{\rm 1}  
\log[k^2x^2 + k^2(y+y_{\rm 00})^2] + \\
\varepsilon S_{\rm 1} \frac{k^2x^2}{1+k^2[x^2+a(y+y_{\rm 00})^2 -x_{\rm 0}^2-b(y_{\rm 0}+y_{\rm 00})^2]}
\, ,
\label{eq:B4}
\eeqa
%
where $k$ is the inverse scale-length, 
$-y_{\rm 00}$ is the vertical coordinate of the singularity in the magnetic field, and 
$a$, $b$, $\varepsilon$, $x_{\rm 0}$, $y_{\rm 0}$, and $y_{\rm 00}$ are dimensionless parameters. 
We set them as 
%
$a=b=0.85$, 
$k = 1$ {\rm Mm}$^{-1}$, 
$x_{\rm 0}=y_{\rm 0}=4$ Mm, 
$y_{\rm 00}=1$ Mm, 
$S_{\rm 1} \approx 11.4$ Gauss Mm, 
%
and hold them constant. 
These loop parameters are chosen to have a small-size loop (averaged radius$\approx$$5$ Mm), 
which significantly simplifies numerical simulations. 
Vectors of magnetic field, resulting from 
Eq.~(\ref{eq:B4}) are illustrated in Fig.~\ref{fig:mag}. 
As a result of the symmetry, the right-hand side of the system is displayed only, and 
the magnetic field vectors consist the arcade with a singularity at ($x=0$, $y=-y_{\rm 00}$) Mm. 

The first term in the right side of Eq.~(\ref{eq:B4}) 
corresponds to 
the potential magnetic arcade, 
in which the magnetic field varies as $1/r$, where $r$ is a radial distance from the axis of symmetry, 
placed 
beneath the photosphere, at 
the location of the singularity. 
Such potential magnetic field does not alter the hydrostatic state of the solar atmosphere 
and the equilibrium mass density and a gas pressure remain equal to $\varrho_{\rm h}(y)$ and $p_{\rm h}(y)$, respectively. 

Since a purely potential arcade 
does not lead to any loop structure, 
we implement 
the small, non-potential 
correction (second) term in Eq.~(\ref{eq:B4}), 
which highlights 
in the body of the magnetic arcade a narrow loop of its radius 
\beq
r_{\rm 0}=\sqrt{x_{\rm 0}^2+y_{\rm 0}^2}\, .
\eeq 
This correction term 
is chosen to have the integral in Eq.~(\ref{eq:B2}) 
evaluated analytically. 
However, the analytical expressions for the equilibrium mass density, $\varrho(x,y)$, and a gas pressure, $p(x,y)$, 
derived by the symbolic package MAXIMA from Eqs.~(\ref{eq:B3}) and (\ref{eq:B2}), 
are 
too long 
to be 
displayed here. 

Within the coronal loop the plasma parameters are 
significantly 
different from those in the ambient corona. 
The correction results in mass density enhancement within the loop. The density ($\varrho$)
within the loop is about twice larger at the loop apex and ten times larger at the loop foot-point 
than the ambient coronal mass density (Fig.~\ref{fig:mag}). 
This loop is about 
four times warmer at its apex and ten times hotter at $x\approx 3$ Mm and $y\approx 3.5$ Mm than the ambient plasma 
(Fig.~\ref{fig:mag}, bottom). 
Below the denser strand-like structure which occupies the top layer of the loop, 
there is the layer of rarefied strand-like plasma at the lower side (Fig.~\ref{fig:mag}, top). 
The whole strands-like structure is about $500$ km wide and the loop is about $15$ Mm long 
with its major radius of about $5$ Mm. It should be noted that loop length, width, and 
major axis etc are the free parameters, and the model can yield the range of loop morphology. We can 
simply mimic the various kinds of coronal loops with different radius of curvature and height, with 
different magnetic field strength, and confined plasma with given density and temperature.
However, here we only choose the small size of the model loop to avoid computationally extensive numerical 
calculations. The major aim of this paper is only to introduce our new coronal loop model with 
a simple example of vertical kink oscillations. Its various applications, and other parametric studies 
will be taken-up in our future 
projects. 
%
\section{Numerical model for vertical kink oscillations
}\label{sec:num_sim_MHD}
To solve 
the 
2D, ideal MHD equations 
numerically, we use 
the FLASH code (Fryxell et al. 2000; Lee \& Deane 2009; Lee
2013), 
in which a third-order unsplit Godunov-type solver with various slope limiters 
and Riemann solvers as well as Adaptive Mesh Refinement (AMR) 
(MacNeice et al. 1999)
are implemented. Among a number of options, we choose the minmod slope limiter and the Roe Riemann 
solver (e.g.,
T\'oth 2000).
%
\begin{figure}[!ht]
	\begin{center}
		\includegraphics[width=7.2cm, angle=0]{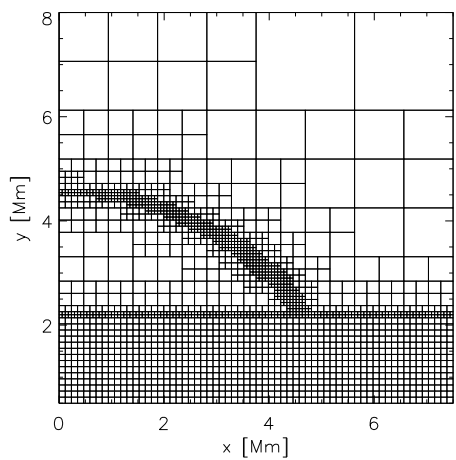}
		\vspace*{-2.25cm}
		\caption{\small Blocks system used in the numerical simulations. 
                        }
		\label{fig:blk}
	\end{center}
\end{figure}
\begin{figure}[!h]
\begin{center}
\includegraphics[scale=0.35, angle=90]{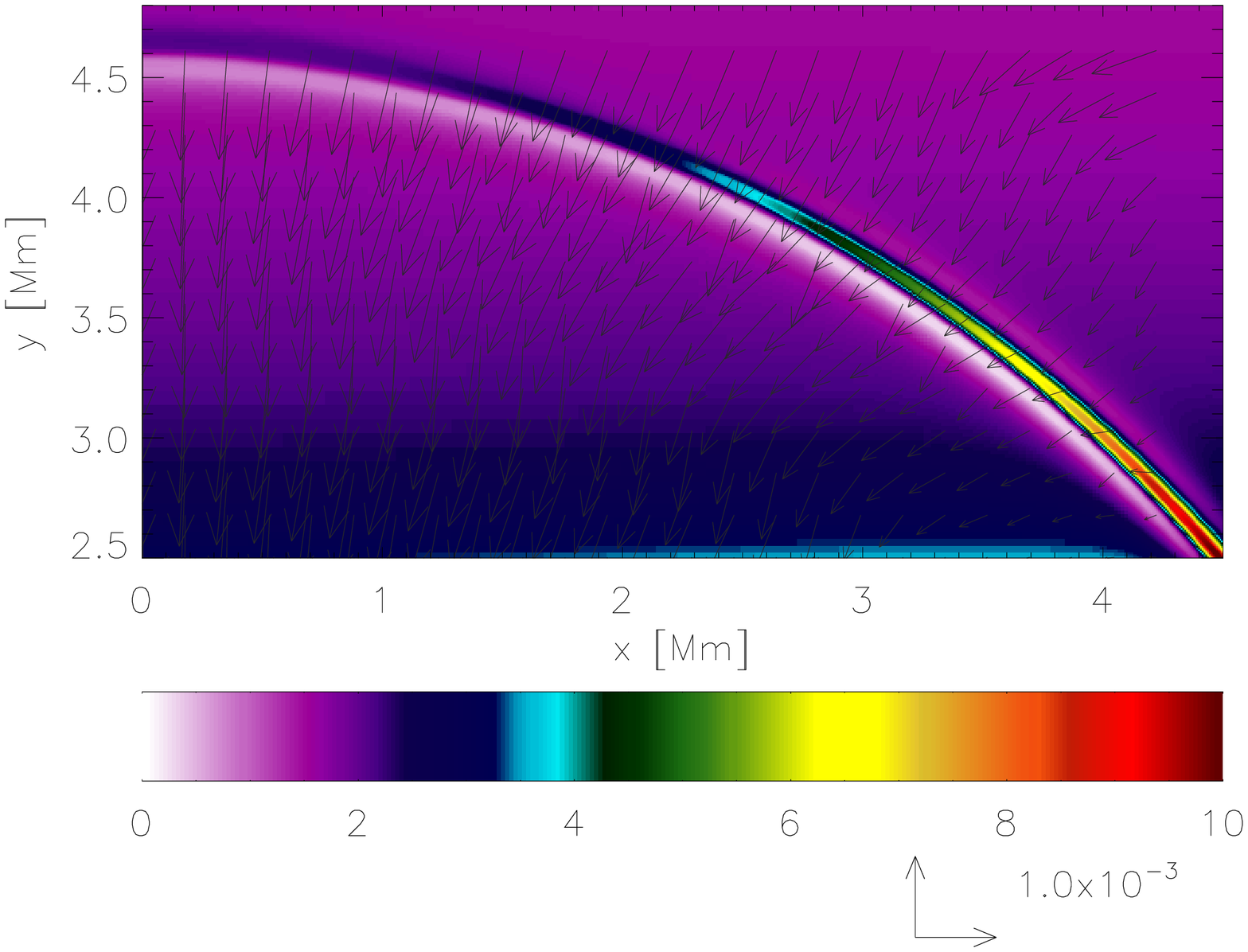}
\includegraphics[scale=0.35, angle=90]{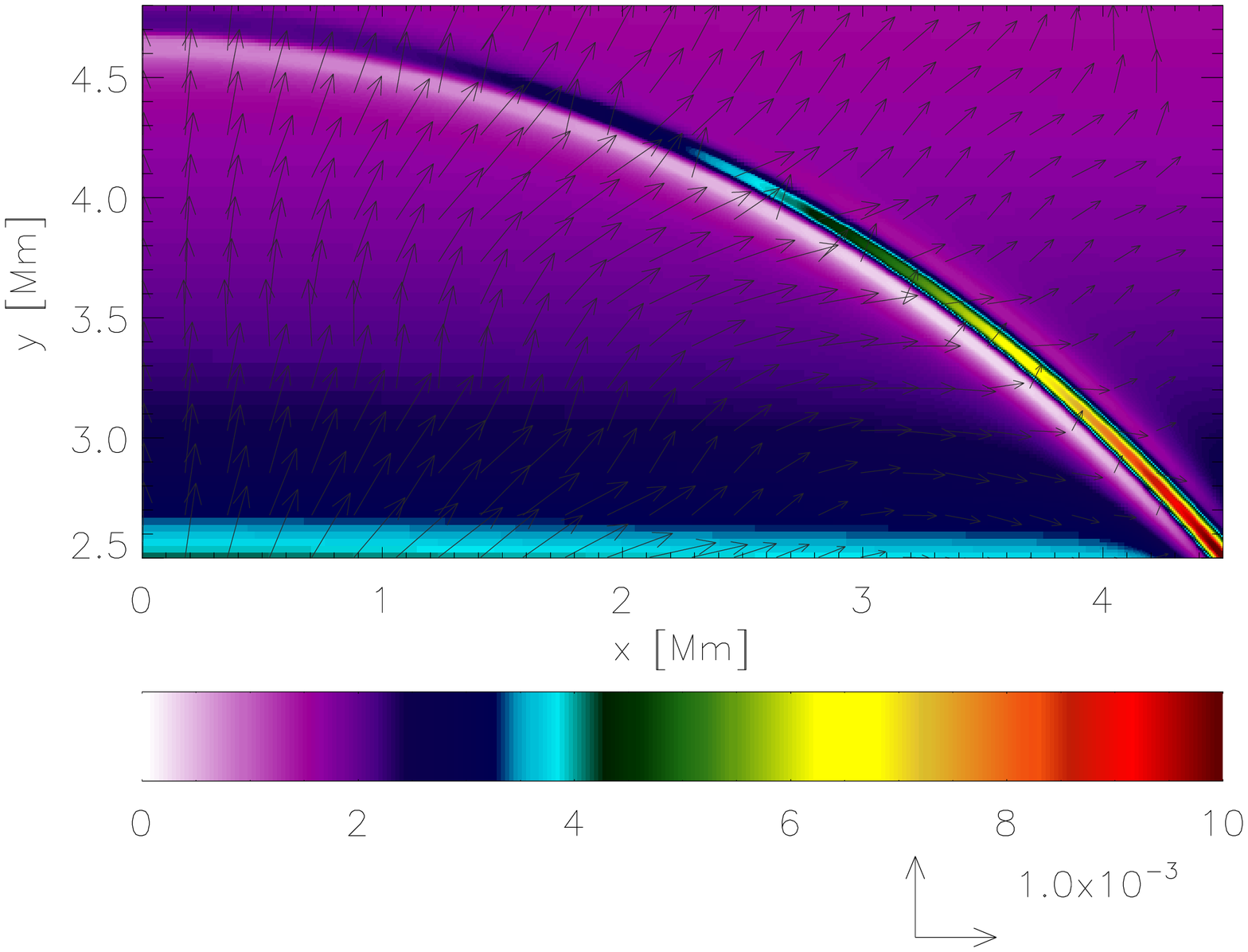}
\vspace{-0.2cm}
\end{center}
\caption{Temporal evolution of $\varrho(x,y)$ at $t=550$ s (top)
and $t=700$ s (bottom). Arrows represent plasma velocity. 
A full-colour version of above figure and movie is available online at www..., Fig4.mpg. 
}
\label{fig:rhofig}
\end{figure}

We set the simulation box as
$(-7.5\, {\rm Mm},7.5\, {\rm Mm}) \times (0.5\, {\rm Mm},8.0\, {\rm Mm})$ 
and impose time-dependent boundary conditions for all plasma quantities
at all four boundaries;  
at these boundaries we set all plasma quantities to their equilibrium values; the only exception
is the bottom boundary, where we additionally place the periodic driver as
%
\beq
\label{eq:Btwist}
V_{\rm y}(x,y,t) = A_{\rm V}\, \exp\left[-\frac{x^{2} + (y-y_{\rm d})^{2}}{w ^{2}} \right] 
                                \sin\left(\frac{2\pi}{P_{\rm d}}t\right) \, ,
\eeq
%
where $A_{\rm V}$ is the amplitude of the driver, $(0, y_{\rm d})$ is its spatial position, 
$w$ denotes its width, and $P_{\rm d}$ is its period. 
We set $A_{\rm V}=3$ km s$^{-1}$, 
$y_{\rm d}=0.5$ Mm, $w = 1$ Mm, $P_{\rm d} = 300$ s, and hold them fixed. 
The driving period $P_{\rm d} = 300$ s corresponds to the average life-time of a solar granule, 
as the granules (together with a random coronal flows, which are although not explored in this paper) 
can be regarded as a real physical driver of decay-less coronal loop oscillations 
(Valery Nakariakov, private communication). 

%
%
\begin{figure*}[!ht]
	\begin{center}
                \mbox{
		\includegraphics[scale=0.4, angle=90]{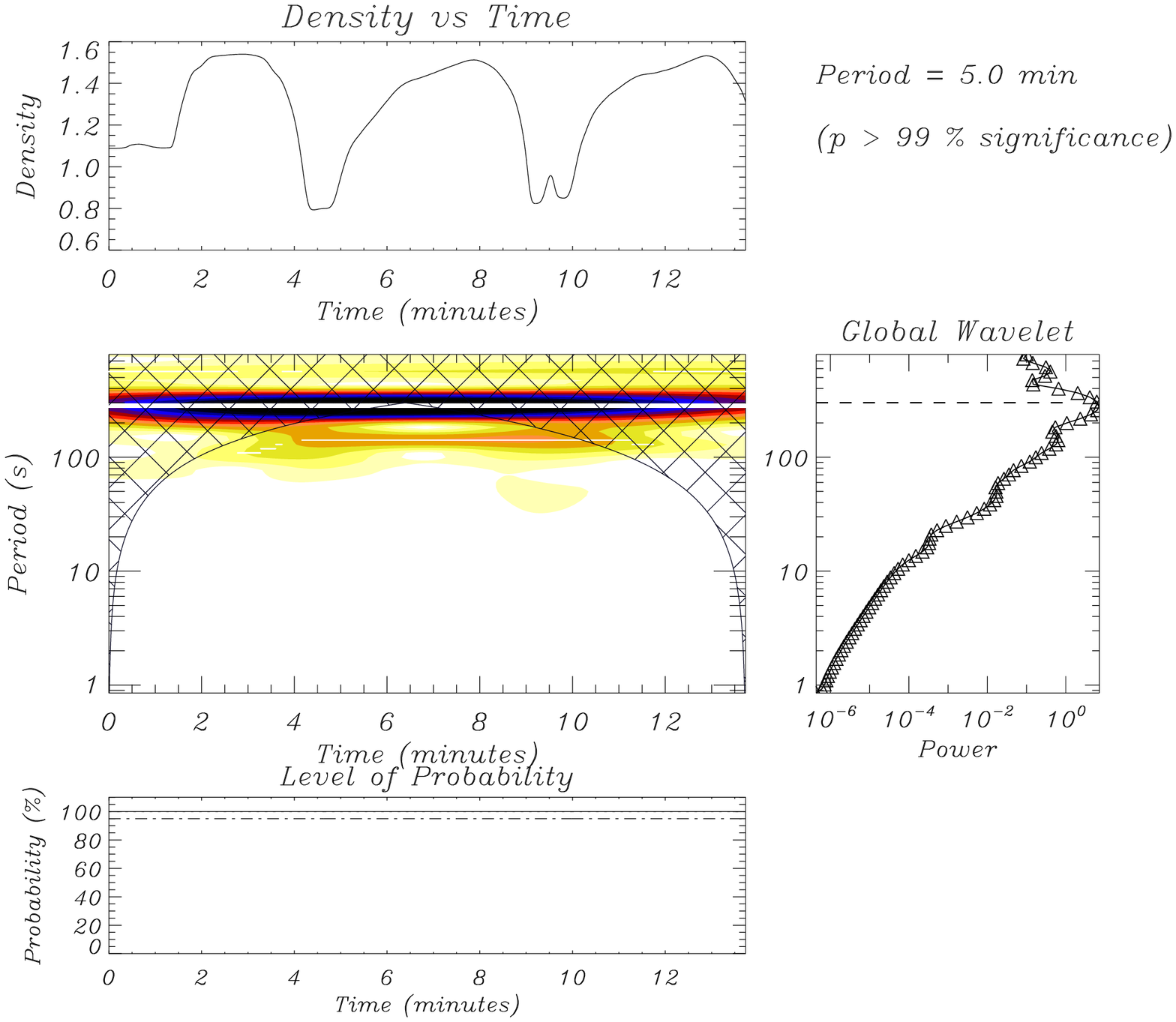}
		\includegraphics[scale=0.4, angle=90]{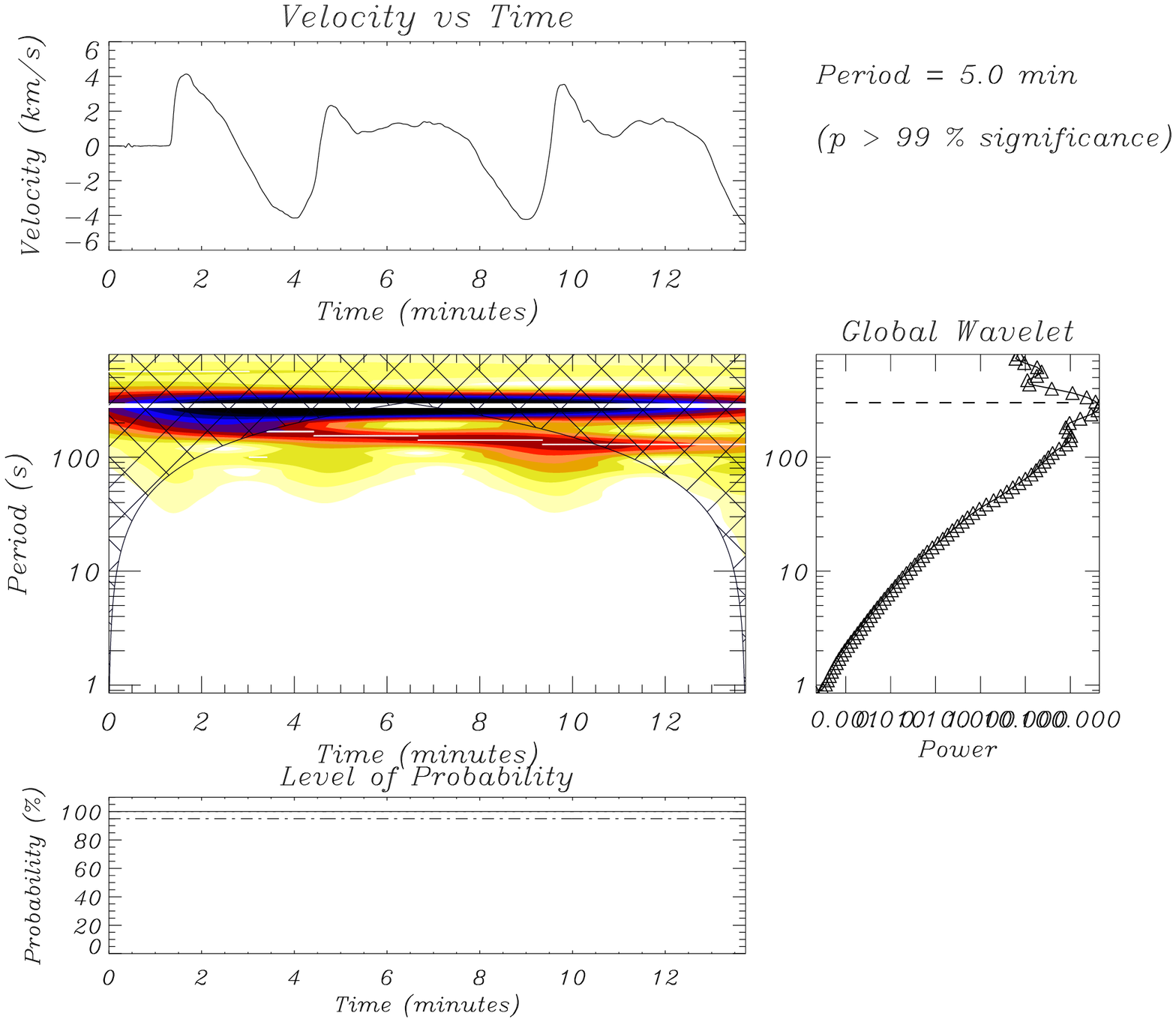}
}
		\caption{\small Time-signatures of $\varrho$ 
(top-left) and $V_{\rm y}$ (top-right) 
collected at $(x=0, y=4.5)$ Mm, and their wavelet spectra.  
                     } 
		\label{fig:wavlet-ts_Vz5}
	\end{center}
\end{figure*}

In our present work, we use an AMR grid with a minimum (maximum) level 
of refinement set to $3$ ($8$). 
We performed 
the grid convergence studies by refining the grid by a factor of two. 
As the numerical results remained essentially same for the grid of maximum block levels $7$ and $8$, 
we adopted the 
latter 
to get the results presented in this paper. 

Note that small-size blocks of numerical grid occupy the altitude upto $y\approx 5$ Mm, 
below the solar transition region and in the neighborhood of the loop (Fig.~\ref{fig:blk}), 
and every numerical block consists of $8\times 8$ identical 
numerical cells. This results in an excellent resolution of steep spatial 
profiles,  
and greatly reduces the numerical diffusion in these regions. 
%
%
%

Figure~\ref{fig:rhofig} shows the spatial profiles of $\varrho(x,y)$ at two time-spans. 
Note that the right-hand part of the simulation region is displayed only. 
As a result of the driver, essentially fast magnetoacoustic-gravity waves are excited in the system. 
The fast magnetoacoustic waves are quasi-isotropic and they propagate upwards across the curved magnetic field lines. 
At $t=550$~s (top) and $t=700$~s (bottom), the  plasma is moving 
downwards (upwards) and the apex of the loop attains approximately its lowest (highest) position of 
$\approx 4.6$ Mm ($\approx 4.7$ Mm), oscillating with the amplitude of about $100$ km. 
It should be noted that the vertical kink oscillations of a curved coronal loop are 
fundamentally different from horizontal kink modes, 
since they are confined to the loop 
plane and can lead to the change of its length.
These oscillations are well seen in 
Fig.~\ref{fig:wavlet-ts_Vz5}
which 
illustrates the time-signature of $\varrho$ (top-left) and $V_{\rm y}$ (top-right) collected at the point $(x=0$, $y=4.5)$ Mm, 
settled just below the apex. 
From this figure we clearly see that the quasi-periodic oscillations are present in the system, and 
the wave-period of these oscillations is equal to the driving period, $P_{\rm d}=300$ s. 
Indeed, Figure~\ref{fig:wavlet-ts_Vz5} (bottom) illustrates wavelet spectra of these time-signatures, 
from which we infer that 
the velocity time-signature is a combination of $5$ min and $3$ min waveperiods, but the $5$ min waveperiod is predominant. 
The periodogram analyses (Scargle 1982) also show the presence of significant ($> 3\sigma$) power peaks 
around $5$ min waveperiod in both mass density as well as velocity time-profiles, which are consistent with the wavelet power spectral analyses. 
The amplitude of these oscillations in $V_{\rm y}$ is about $1.1$ km s$^{-1}$ (Fig.~\ref{fig:wavlet-ts_Vz5}, top-right).
It should be noted that the phase-difference between velocity and density variations with time
is $\pi$/4 (quarter period), which is a typical property of the fundamental kink oscillations excited in a loop. 
%
\section{Summary} 
\label{sec:summary}
In the present paper, we have presented for the first time the analytical and numerical models of a solar coronal loop, 
which is embedded in a gravitationally stratified solar corona. 
These models are 
based on the analytical models of Solov’ev (2010).
The Kra\'skiewicz et al. (2014) and Ku\'zma
et al. (2014)
report 
can also be referred for the detailed mathematical formulation of the analytical and numerical models. 
Using these 
models, 
we performed 
the 2D numerical simulations 
of the vertical kink oscillations of this loop excited by the forced periodic driver 
that acts at the top of the photosphere, centrally below the apex of the loop. 

The numerical simulations adapt the realistic model of the hydrostatic solar atmosphere in the FLASH code 
and a slightly modified potential magnetic field. 
Our model exhibits the formation of the quasi-periodic vertical oscillations 
of their wave-period equal to the driving wave-period of $300$ s and 
their velocity amplitude is about $1.2$ km s$^{-1}$, while the loop apex oscillates with its 
amplitude of $100$ km. These values match the recent observational findings of
Nistic\'o et al. (2013) and Anfinogentov et al. (2013). 

We note here that 
the implemented driver 
mimics a downdraft associated with a solar granule as 
the driving period 
is set to the 
life-time of a granule, that is $300$ s, 
and its amplitude is $3$ km s$^{-1}$. However, the latter value seems to be larger by a factor of about $3$ than 
the downdraft speed. We have verified by numerical experiments that a lower amplitude of the driver 
resulted in less pronounced oscillations (not shown). 
Moreover, to simplify numerical simulations we have chosen the coronal loop of only about $15$ Mm long, 
which is at least an order of magnitude too short than a typical coronal loop. 
In the latter case, the fast magnetoacoustic waves would experience more spatial spreading while 
propagating from the launching place upwards 
towards 
the typical loop apex and covering a long distance. 
As a result of that this apex would be affected by less energetic signal and it would experience 
lower amplitude oscillations. In this case a larger amplitude of the periodic driver would be required 
or a driver can be set higher-up, somewhere in the solar corona, modelling its random velocity field. 
Qualitatively, we can state that the oscillation amplitude should
decline with the loop length (L) and it should grow with the amplitude
of the forced driver (A$_{v}$). Such parametric studies would be important for
impulsively excited waves with the non-forced drivers, which will be devoted to our future study using 
the newly developed coronal loop model.


%
%
\begin{acknowledgements} 
We thank the referee for his/her valuable comments which improved 
the manuscript considerably. 
AKS thanks Prof. K. Murawski and UMCS, Lublin for providing a visit fund during September-October 2014 during which He contributed to the project. 
The work has also been supported by a Marie Curie International 
Research Staff Exchange Scheme Fellowship within the 7th European
Community Framework Program. 
The software used in this work was in part developed by the 
DOE-supported ASCI/Alliance Center for Astrophysical Thermonuclear
Flashes at the University 
of Chicago.  The visualizations of the simulation variables have been
carried out 
using the IDL (Interactive Data Language) software package.
\end{acknowledgements}
%


\begin{thebibliography}{qqq}
\bibitem[Anfinogentov et al.(2013)]{2013A&A...560A.107A} Anfinogentov, S., Nistic{\`o}, G., \& Nakariakov, V.~M.\ 2013, \aap, 560, AA107
\bibitem[Antolin et al.(2014)]{2014ApJ...787L..22A} Antolin, P., Yokoyama, T., \& Van Doorsselaere, T.\ 2014, \apjl, 787, LL22
\bibitem[Aschwanden et al.(1999)]{1999ApJ...520..880A} Aschwanden, M.~J., Fletcher, L., Schrijver, C.~J., \& Alexander, D.\ 1999, \apj, 520, 880
\bibitem[Aschwanden et al.(2000)]{2000ApJ...541.1059A} Aschwanden, M.~J., Nightingale, R.~W., \& Alexander, D.\ 2000, \apj, 541, 1059
\bibitem[Aschwanden \& Schrijver(2011)]{2011ApJ...736..102A} Aschwanden, M.~J., \& Schrijver, C.~J.\ 2011, \apj, 736, 102
\bibitem[Avrett \& Loeser(2008)]{2008ApJS..175..229A} Avrett, E.~H., \& Loeser, R.\ 2008, \apjs, 175, 229
\bibitem[De Moortel \& Nakariakov(2012)]{2012RSPTA.370.3193D} De Moortel, I., \& Nakariakov, V.~M.\ 2012, Royal Society of London Philosophical Transactions Series A, 370, 3193
\bibitem{Fryxell2000} Fryxell, B., Olson, K., Ricker, P., et al.\ 2000, \apjs, 131, 273
\bibitem[Gruszecki et al.(2006)]{2006A&A...460..887G} Gruszecki, M., Murawski, K., Selwa, M., \& Ofman, L.\ 2006, \aap, 460, 887
\bibitem{Kraskiewicz2014} Kr\'askiewicz, J., Murawski, K., Solov'ev, A. \& Srivastava, A.K.\ 2014, Sol. Phys., submitted.
\bibitem{Kuzma2014} Ku\'zma, B., Murawski, K., Solov'ev, A. \ 2014, \aap, submitted.
\bibitem{Lee2013} Lee, D.\ 2013, Journal of Computational Physics, 243, 269 
\bibitem{Lee2009} Lee, D. \& Deane, A. E.\ 2009, Journal of Computational Physics, 228, 952 
\bibitem{Low1975} Low, B.~C.\ 1975, \apj, 197, 251
\bibitem[Luna et al.(2010)]{2010ApJ...716.1371L} Luna, M., Terradas, J., Oliver, R., \& Ballester, J.~L.\ 2010, \apj, 716, 1371
\bibitem{MacNeice1999} MacNeice, P., Spicer, D.~S., \& Antiochos, S.\ 1999, 8th SOHO Workshop: Plasma Dynamics and Diagnostics in the Solar Transition Region and Corona, 446, 457
\bibitem[Nistic{\`o} et al.(2013)]{2013A&A...552A..57N} Nistic{\`o}, G., Nakariakov, V.~M., \& Verwichte, E.\ 2013, \aap, 552, AA57
\bibitem[Ofman(2009)]{2009ApJ...694..502O} Ofman, L.\ 2009, \apj, 694, 502 
\bibitem[Ofman \& Wang(2008)]{2008A&A...482L...9O} Ofman, L., \& Wang, T.~J.\ 2008, \aap, 482, L9
\bibitem[Pascoe(2014)]{2014RAA....14..805P} Pascoe, D.~J.\ 2014, Research in Astronomy and Astrophysics, 14, 805
\bibitem{Priest1982} Priest, E.~R.\ 1982, Dordrecht, Holland ; Boston : D.~Reidel Pub.~Co.~; Hingham,,
\bibitem[Scargle(1982)]{1982ApJ...263..835S} Scargle, J.~D.\ 1982, \apj, 263, 835 
\bibitem[Selwa et al.(2011)]{2011ApJ...728...87S} Selwa, M., Solanki, S.~K., \& Ofman, L.\ 2011, \apj, 728, 87 
\bibitem[Solov'ev(2010)]{2010ARep...54...86S} Solov'ev, A.~A.\ 2010, Astronomy Reports, 54, 86 
\bibitem[Srivastava \& Goossens(2013)]{2013ApJ...777...17S} Srivastava, A.~K., \& Goossens, M.\ 2013, \apj, 777, 17
\bibitem{Toth2000} T\'oth, G.\ 2000, Journal of Computational Physics, 161, 605
\bibitem[Verwichte et al.(2009)]{2009ApJ...698..397V} Verwichte, E., Aschwanden, M.~J., Van Doorsselaere, T., Foullon, C., 
\& Nakariakov, V.~M.\ 2009, \apj, 698, 397
\bibitem[Wang \& Solanki(2004)]{2004A&A...421L..33W} Wang, T.~J., \& Solanki, S.~K.\ 2004, \aap, 421, L33 
\bibitem[Wang et al.(2008)]{2008A&A...489.1307W} Wang, T.~J., Solanki, S.~K., \& Selwa, M.\ 2008, \aap, 489, 1307
\bibitem[White et al.(2012)]{2012A&A...545A.129W} White, R.~S., Verwichte, E., \& Foullon, C.\ 2012, \aap, 545, AA129 
\end{thebibliography}
\end{document}